\documentclass[a4paper,prl,floatfix,twocolumn,superscriptaddress,showpacs]{revtex4}
\usepackage[USenglish]{babel}
\usepackage{amsmath,amssymb,amsfonts}
\usepackage{color}
\usepackage{graphics,graphicx}
\usepackage{color}

\usepackage[colorlinks=true,linkcolor=blue,citecolor=red]{hyperref}

\def\bcen{\begin{center}}
\def\ecen{\end{center}}

  \def\ie{\mbox{\it i.e.\ }}
\def\=={\equiv}

\def\qed{\raise1pt\hbox{\vrule height5pt width5pt depth0pt}}

\def\cG0{{\cal G}_0} 
\def\cG{{\cal G}}

  \def\Im{\mbox{Im}}
\def\ie{\hbox{\it i.e.\ }} 
\def\ie{\mbox{\it i.e.\ }} \def\=={\equiv}
 
\def\Im{{\rm Im}}  
 \def\ep0{\epsilon_{p}} \def\ed0{\epsilon_{f}}

\begin{document}

\author{A. Amaricci}
\affiliation{Democritos National Simulation Center, 
Consiglio Nazionale delle Ricerche, 
Istituto Officina dei Materiali (IOM) and 
Scuola Internazionale Superiore di Studi Avanzati (SISSA), 
Via Bonomea 265, 34136 Trieste, Italy}

\author{A. Privitera}
\affiliation{Democritos National Simulation Center, 
Consiglio Nazionale delle Ricerche, 
Istituto Officina dei Materiali (IOM) and 
Scuola Internazionale Superiore di Studi Avanzati (SISSA), 
Via Bonomea 265, 34136 Trieste, Italy}

\author{M.Capone} 
\affiliation{Democritos National Simulation Center, 
Consiglio Nazionale delle Ricerche, 
Istituto Officina dei Materiali (IOM) and 
Scuola Internazionale Superiore di Studi Avanzati (SISSA), 
Via Bonomea 265, 34136 Trieste, Italy}

\title{Inhomogeneous BCS-BEC crossover for trapped cold atoms in optical lattices}
\begin{abstract}

The BCS-BEC crossover in a lattice is a powerful paradigm to
understand how a superconductor deviates from the
Bardeen-Cooper-Schrieffer physics as the attractive interaction increases. 
Optical lattices loaded with binary mixtures of cold atoms allow to
address it experimentally in a clean and controlled way.
We show that, however, the possibility to study this phenomenon in actual cold-atoms
experiments  is limited by
the effect of the trapping potential. Real-space Dynamical Mean-Field
Theory calculations show indeed that interactions and the
confining potential conspire to pack the fermions in the center of the
trap, which approaches a band insulator when the attraction become
sizeable. We show how this physics reflects in several observables, and
we propose an alternative strategy to disentangle the effect of the
harmonic  potential and measure 
the intrinsic properties resulting from the interaction strength.
\end{abstract}
\pacs{71.10.Fd, 03.75.Ss, 05.30.Fk, 67.85.Lm}
\maketitle

The experimental advances in handling and probing cold atoms in
optical lattices open a new path towards the understanding of
popular condensed-matter lattice models\cite{Bloch2012NP}. 
While the repulsive Fermi-Hubbard model and its Mott insulating phase\cite{Jordens2008N,Jordens2010PRL}
are the first natural goal because of their relation with high-temperature
superconductivity, the experimental
realization\cite{hackermueller2010S} of the attractive Fermi-Hubbard model
(AHM)  is an equally sensible target. 
The quantum simulation of the AHM has at least a twofold motivation: besides
its direct significance as an idealized description of actual
superconductors, it has been proposed  as a simpler path to
investigate the repulsive 
model\cite{Ho2009PRA} exploiting an exact mapping between the
two models.

At low temperature the AHM describes a superfluid (SF) state, whose
properties evolve continuously from a weak-coupling Bardeen-Cooper Schrieffer (BCS)
regime to a Bose-Einstein condensation (BEC) of preformed pairs as the
attractive interaction is increased. The lattice counterpart of the
BCS-BEC crossover\cite{nozieres1985} has been proposed as an
effective  description of high-temperature superconductors, and it displays 
significant differences with the crossover of dilute Fermi 
gases\cite{Zwerger} including a pronounced maximum for intermediate 
pairing strength of the critical temperature, which vanishes as $1/U$ 
for large attraction and a characteristic dependence on the lattice density,
\ie the number of fermions ($N$) per lattice site $n = N/N_s$.

The description of the lattice BCS-BEC crossover requires
non-perturbative approaches, among which Dynamical mean-field
theory (DMFT)\cite{Georges1996RMP} can be particularly useful, 
as it correctly reproduces the exact solution both in the weak- and in the strong-coupling 
limit\cite{Garg2005PRB,Toschi2005NJOP} as
well as the evolution of the normal state from which superfluidity
establishes\cite{Keller2001PRL,Capone2002PRL}. DMFT also recovers the
familiar  BCS-BEC crossover for a  Fermi gas in the 
dilute limit\cite{Privitera2010PRB,Privitera2012PRA}. 

However, DMFT enforces translational symmetry, which is clearly broken by the
harmonic potential which traps the fermions in cold-atoms
experiments. 
This requires to use an extension of DMFT, the real-space DMFT
in order to take into account the inhomogeneity of the system and 
to investigate the effect of the trap on the BCS-BEC crossover. 
The same method, with a different impurity solver 
(see below), has been used in Ref.\onlinecite{Koga2009PRA} to identify
a coexistence of SF and density-wave. While our focus is different, we
mention that we did not observe a tendency to density ordering, in
agreement  with the Quantum Monte Carlo results of Ref.\onlinecite{Assmann2012PRB}.

Our zero-temperature calculations show that increasing the attraction
strength leads to a compression of the cloud, with a central region
populated by two fermions of opposite spin per lattice site, as in a band
insulating state, leading to a packed cloud with reduced pairing
amplitude.  
This collapse as a function of the interaction prevents us from reaching 
the actual BEC regime of the AHM, where local pairs are formed, but
they do not coalesce in the same region of space. 
Indeed, the anomalous expansion of the cloud observed
in experiments\cite{hackermueller2010S} does not overcome this limitation, as
it is essentially due to adiabatic heating\cite{Schmidt2013PRL}, an effect
which introduces a further obstacle to the observation of the BCS-BEC
crossover by effectively increasing the temperature at fixed
entropy. 
We characterize the hidden crossover with observables which are
accessible in the current cold-atoms experiments, like the momentum
distribution function and the single-particle spectral functions. 
In addition, we propose a simple way to reduce the impact of the 
cloud compression and unveil the ``homogeneous'' BCS-BEC crossover 
compensating the effect of the inhomogeneous potential.

\begin{figure}
 \includegraphics[width=0.4925\linewidth]{fig1a.eps}
 \includegraphics[width=0.4\linewidth]{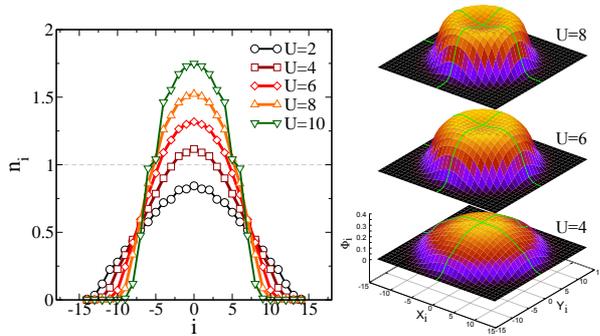} 
 \caption{(Color online) Left panel: local density $n_i$ profiles
   along the x-axis of the lattice ($y_i=0$) for $N=200$ fermions,
   $V_0=0.03$,  and increasing attraction $U$. Data are for a lattice
   of $N_s=$29x29 sites. Right panel: evolution
   of the corresponding superfluid amplitude $\phi_i$ surface and
   profiles for $U=4$ (bottom),  $U=6$ (center) and $U=8$ (top). 
}\label{fig1} 
\end{figure}

In all the calculations we consider an attractive Fermi-Hubbard model
on a  two-dimensional square optical lattice 
and placed in an external harmonic potential. The Hamiltonian reads:
\begin{equation}
 {\cal H}=-t\sum_{\langle i,j\rangle \sigma}
 c^\dagger_{i\sigma}c_{j\sigma} 
- U\sum_{i} n_{i\uparrow} n_{i\downarrow} + 
\sum_{i\sigma} V_i n_{i\sigma}
\label{Ham}
\end{equation}
where $t$ is the hopping parameter between neighboring sites, which we
set as the energy unit.
The second term describes the local attractive interaction between fermions.
Finally, the last term $V_i = \frac{V_0}{2}(r_i/a)^2$ is the harmonic
trapping potential, that we assumed with spherical symmetry, $a$ is the
the lattice spacing and $r_i$ is the distance of the site $i$ 
from the trap center.

We solve Eq.~(\ref{Ham}) on a lattice of $N_s$ sites, using
real-space DMFT\cite{Helmes2008PRL,Andrade2009PRL,Miranda2013}, 
an extension of DMFT\cite{Georges1996RMP} introduced to treat 
inhomogeneous system.
The key approximation is to assume a local, albeit site-dependent, self-energy 
matrix $\hat{\Sigma}_{ij}=\delta_{ij}\hat{\Sigma}_i$. Each local
self-energy is obtained by solving an impurity problem defined by a
site-dependent  bath $\hat{\cal G}_{0i}^{-1}$, which is determined
self-consistently by requiring that the single-particle Green function
$\hat{G}_{i}$ of each local impurity model coincides with the
corresponding  diagonal term of  $\hat{G}^{-1}=\hat{G}_0^{-1}-\hat{\Sigma}$, where
$(\hat{G}_0^{-1})_{ij}=\delta_{ij}[\omega^+ - (V_i-\mu)]
-\hat{t}_{ij}$ is the non-interacting Green's function and
$\hat{t}_{ij}$ is the lattice tight-binding matrix.
In order to deal with superfluid phase, we recast the method in the 
Nambu spinor formalism\cite{Koga2009PRA}, introducing anomalous (pair) 
Green's functions and self-energy components 
$\hat{F}_{i}$ and $\hat{S}_{i}$, respectively. 

\begin{figure}
 \includegraphics[width=0.65\linewidth]{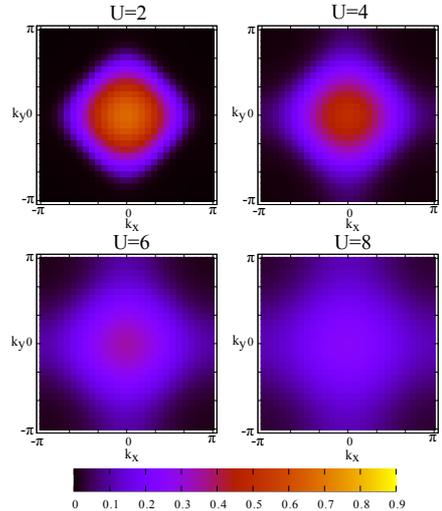}
 \caption{(Color online) Evolution of the momentum distribution $n_k$
   as a function of the increasing attraction $U$. Data are obtained
   for a lattice of $N_s=$29x29 sites. The other model parameters as in Fig.~\ref{fig1} 
}\label{fig2} 
\end{figure}
The number of independent impurity models is reduced by the lattice
$C_{4v}$ symmetry. The solution of the impurity problems is obtained using the 
Iterated Perturbation Theory solver\cite{Kajueter1996PRB,Georges1996RMP}, 
extended to deal with superconducting formalism\cite{Garg2005PRB}. 
This method provides an accurate and computationally cheap solver
which gives direct access to dynamical properties including the local spectral functions 
$\rho_i(\omega)=-\Im G_{i}(\omega)/\pi$ at the site $i$ and 
hence to the local spectral gap $E^g_i$. 
This information can be experimentally accessed by a spectroscopic
technique able  to probe the local value of the gap (see e.g. \cite{Gericke2008NP} for a
cold-atom analogue of the scanning tunneling microscopy used in
condensed matter).  
We shall compare our calculations with local-density
approximation (LDA) results where the local observables on each site 
are those of a homogeneous system with chemical potential
$\mu_i=\mu-V_i$.

\begin{figure*}[htb]
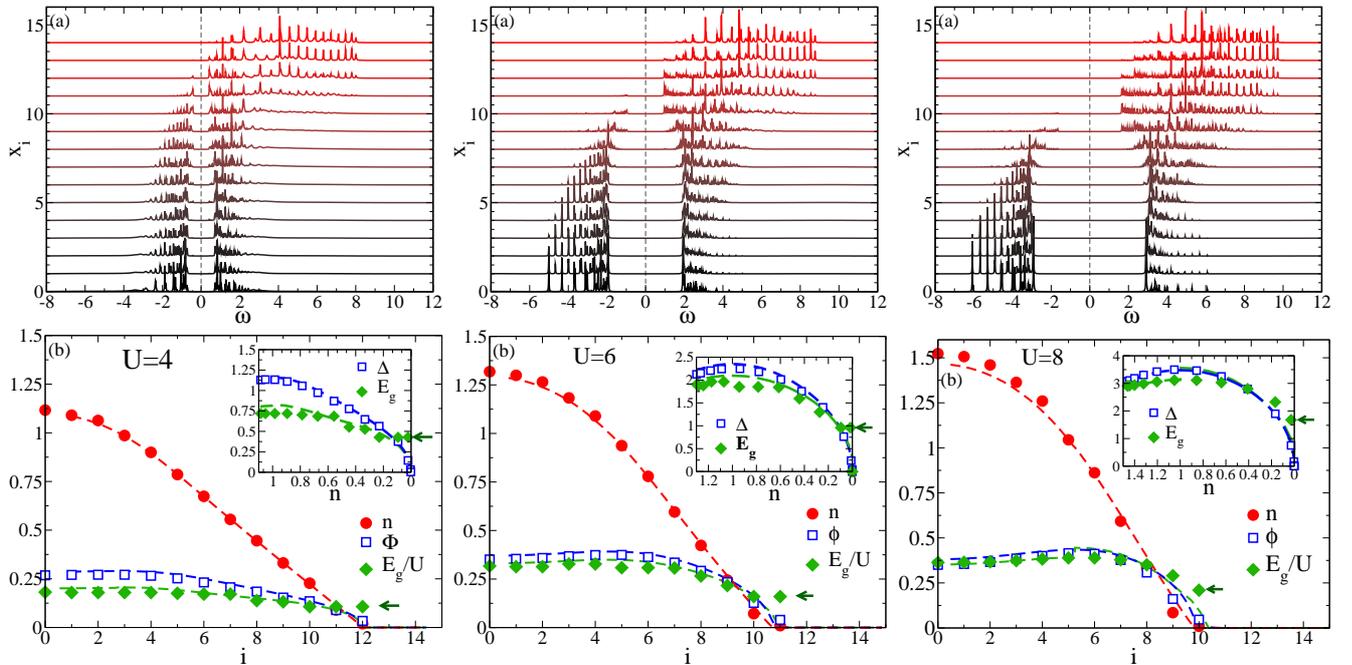

\begin{center}
\includegraphics[width=0.32\textwidth]{fig3_panelb_U4_Xaxis.eps}
\includegraphics[width=0.32\textwidth]{fig3_panelb_U6_Xaxis.eps}
\includegraphics[width=0.32\textwidth]{fig3_panelb_U8_Xaxis.eps}

\includegraphics[width=0.32\textwidth]{fig3_panela_U4.eps}
\includegraphics[width=0.32\textwidth]{fig3_panela_U6.eps}
\includegraphics[width=0.32\textwidth]{fig3_panela_U8.eps}
\caption{(Color online) Spectral functions (a) and energy gap (b) 
  evolution of the trapped system along the $y_i=0$ axis for $U=4$ (left), $U=6$
  (center) and $U=8$ (right). Other parameters are: $V_0=0.03$,
  $N=200$ and system size is $N_s=$29x29 sites. The top panels report the
  distribution of the  density $n_i$, the  amplitude $\phi_i$
  and the gap $E^g_i$. The insets show the behavior of the order parameter
  $\Delta_i$ and  gap $E^g_i$ as a function of the density. The
  arrows indicate the large discontinuity at the border of the cloud. 
}
\label{fig3}
\end{center}
\end{figure*}

We briefly recall the main properties of the SF phase for the
homogeneous Hubbard model. The modulus of the superfluid order parameter
$\phi = 1/N_s\sum_i \phi_i = 1/N_s\sum_i \langle c_{i\uparrow}
c_{i\downarrow}\rangle$ and the spectral gap $E^g$ monotonically
increases as a function of $U$, 
while the critical temperature decreases for intermediate and
large $U$ because the large pairing strength locks the fermions in on-site
pairs, which are strongly bound, but they move only through virtual
processes of order $t^2/U$ (a small number if $U \gg t$), making it
harder  and harder to establish phase coherence over the whole system,
a necessary condition for a SF state. 
As a consequence, the critical temperature in this regime is
controlled by the superfluid stiffness, in turn proportional to
$t^2/U$ and decreases rapidly, as opposed to the weak-coupling regime,
where the standard result $T_c \propto \phi$ is recovered. The BEC
side of the crossover is characterized also by a kinetic-energy gain
which stabilizes the SF state, in contrast with the BCS theory, where
a potential-energy gain leads to the SF\cite{Toschi2005NJOP}. 
Pairing without phase coherence results, in strong-coupling, to
a normal state with a pseudogap in the spectrum.
The lattice periodicity also  introduces a peculiar dependence on the
density.   $\phi$ is non monotonic as a function of the lattice
filling, with a maximum at half-filling $n=1$ and a vanishing value for
empty and completely filled lattice. 

In Fig.~\ref{fig1} we show the evolution of the 
density profile $n_i = \sum_\sigma \langle c^\dagger_{i\sigma} c_{i\sigma} \rangle$
and of the local pairing amplitude $\phi_i$ for increasing $U$ along
the $y_i=0$ 
for $N\! =\!200$ fermions on a lattice of $N_s=$29x29 sites.
As pointed out in Ref.~\onlinecite{Koga2009PRA}, by defining 
a typical radius $r_c$ such that $\frac{V_0}{2}(r_c/a)^2=t$ and rescaling 
the density profiles in units of $r_c$, the results for fixed $\mu$ and increasing $r_c$, 
\ie increasing $N$ and decreasing $V_0$, nicely collapse on the same curve. 
Thus our results are directly relevant for current experiments in ultracold 
gases as they can be easily extrapolated to actual system size and
number of particles.

The density profiles for moderate and large $U$ show that the confining potential and the 
interaction concur in pushing the fermions towards the trap center and in 
squeezing the cloud size. 
This effect is clearly triggered by the presence of the harmonic
potential which favors a higher occupation of the
central region. In the presence of an attractive interaction, this tendency 
is further enhanced by the energy gain associated to doubly occupied sites. 
This leads, as the interaction grows, to a packing of the central region, in
which most of the fermions are confined, 
which approaches a local density of $n=2$ (as for a band insulator), 
giving rise to a more compact
cloud with sharper boundaries with respect to a repulsive case, in which the
interaction spreads the fermions in space.

The local superfluid amplitude $\phi_i$, shown in the
right side of Fig.\ref{fig1} , has a non-trivial evolution. For weak
interaction  $\phi_i$ is maximum at the trap center, and decreases
monotonously  moving towards the edges of the condensate. Increasing
the interaction,  for $U=6t$ the maximum at the center turns into a
minimum  while a shallow maximum develops at a distance from the
center.  By further increasing the interaction the maximum moves at
larger distances,  while the whole pairing profile decreases.

This behavior can be traced back -in a LDA scheme- to nonmonomotic
behavior as a function of filling, which is symmetric around a maximum
at half-filling. Increasing the local density beyond half-filling is 
therefore expected to lead to a decrease of
$\phi$. 
For our number of electrons, which would correspond to a density $n \simeq 0.238$ in a homogeneous system, at weak coupling the cloud compression 
due to trap and  interaction is not strong enough to raise the local density
$n_0$ in the trap center above 1. 
In this case $\phi_i$ is 
maximum in the trap center and decreases monotonously as a function of the 
distance from the trap center. At large $U$ instead the cloud
compression becomes strong enough to have $n_0 > 1$ and, although the
attraction  is larger, the order 
parameter in the trap center is suppressed, and the SF order amplitude
acquires a ring shape, with a maximum amplitude around the line where the local
density crosses $n_i =1$. 

It is important to notice that, besides the peculiar spatial pattern
of the pairing amplitude, the collapse of the fermionic cloud
significantly reduces the whole superfluid properties with respect to
a homogeneous system with the same interaction strength and number of
fermions. This is associated to the proliferation of empty and doubly
occupied sites, configurations that share a vanishing pairing amplitude. 
Therefore the BCS-BEC crossover we would observe
in a homogeneous system is hidden by this effect, which starts already
for intermediate coupling. 

The same crossover is reflected in the momentum
distribution function $n_k$ (see Fig.~\ref{fig2}), which is easily accessible in time-of-flight
measurements. Once again, the DMFT results spotlight a rapid evolution
from a BCS regime, characterized by ballistic expansion of the
fermions to an intermediate coupling in which most fermions are
gathered in the center of the trap. The evolution of $n_k$ as a
function of $U$ shows indeed how the remnant Fermi surface abruptly turns into a broad
distribution, characteristic of localized incoherent particles.

\begin{figure}
\begin{center}
\includegraphics[width=0.4\textwidth]{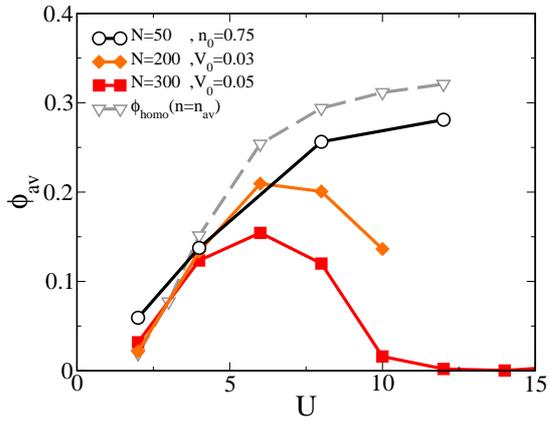}
\caption{(Color online) Superfluid amplitude $\phi$ as a function of
  interaction $U$. Data are obtained using different protocols for the
  BCS-BEC crossover using a total number of particles $N=50$ (open
  circles)  $N=200$ (filled triangles) and $N=300$ (filled squares)
  respectively  fixing the density at the center to $n_0=0.75$ or
  fixing  the harmonic potential strenght to $V_0=0.03$ (triangles)
  and  $V_0=0.05$ (squares). Dashed line indicates the homogeneous
  solution $\phi_{homo}$ at average density $n_{av}$ for $N=50$ and $n_0=0.75$.
}
\label{fig4}
\end{center}
\end{figure}

We now discuss how this physics reflects in
the single-particle spectra and in the momentum distribution function,
before propoposing a simple way to unveil the properties of the
homogeneous crossover. In the top panels of Fig.~\ref{fig3} we show the
evolution of the local single-particle spectral 
function $\rho_i(\omega)=-1/\pi \Im G_{i}(\omega^+)$ along a
cut parallel to the x-axis from the trap center (bottom) to the edge
(top) for three different values of $U$. Notice that the discrete
nature  of the spectral function is a genuine feature due
to the finite lattice and the trapping potential and it does not result
from the DMFT treatment or from our solution method. 

The main changes in the spectral functions as we move from the center to the
boundary of the trap are associated to the change in local
density. Interestingly, the energy gap $E^g_i$ appears more uniform
than the whole spectral function. 
In the lower panels of Fig.~\ref{fig3} we report $E^g_i$ as a
function of the lattice position, together with 
the corresponding values of $n_i$, ${\phi}_i$. For the sake of
comparison, $E^g_i$ is divided by $U$, so that it can be more closely
compared with $\phi_i$ (in the BCS regime $E^g = U\phi$). 
LDA results are shown for comparison as dashed lines. 

Even if the global change in the curves going from the center to the
edge of the trap may suggest that the DMFT results are well reproduced
by LDA, significant deviations appear in the most delicate border
region (notice that the center of the trap hosts an essentially
trivial state). 
Interestingly, the spectral gap shows the most significant deviations
with respect to LDA. $E^g_i$ remains indeed essentially uniform in
space also in the proximity of the cloud edge, while $\phi_i$ vanishes
as predicted by LDA. This leads to a strong deviation from the BCS
proportionality between the two observables. As a matter of fact, the
boundary of the cloud behaves like a phase-disordered superconductor with a
finite spectral gap which is not accompanied by an actual SF order
parameter. A similar behavior is indeed observed in Ref.~\onlinecite{Sacepe2011NP} 
in the context of chemically disordered superconductors. 

Our DMFT solution of the AHM in a trapping potential prompts that, in
order to reveal the full BCS-BEC crossover, a more careful ad-hoc
protocol has to be used. In particular one needs to compensate the
cloud compression due to the increased interaction and keep the
density as uniform as possible and, most importantly, independent on $U$. 
The simplest knob we can use to this end is the strength of the
trapping potential. When $U$ increases, we can decrease $V_0$ and compensate
for the cloud compression. As a matter of fact, it turns out that a
suitable change of $V_0$ is sufficient to reproduce an essentially
constant density pattern for a wide range of $U$. This compensation
protocol avoids the collapse of the cloud and allows for a sensible
comparison between different values of $U$. 

In Fig.~\ref{fig4} we show the performance of this
compensation protocol. We perform calculations for different values of
$U$, choosing $V_0$ in order to keep constant the density at the trap
center $n_0 = 0.75$. 
In order to compare results obtained with different protocols and
potential widths, we estimated the average value of observables 
by averaging over sites with local occupancy larger than a small
threshold $n_i > 0.001$. 
Even with this simple requirement, also the average density in the
cloud is essentially constant as $U$ goes from 2 to 12. 
The success of this choise in revealing the properties of the BCS-BEC
crossover is testified by the main panel of Fig.~\ref{fig4}, where the
average pairing amplitude $\phi_{av}$ for both the straightforward
calculations at fixed $V_0$ and for fixed $n_0$. 
It is apparent that calculations at fixed $V_0$ fail in describing the monotonic increase
of $\phi_{av}$ as the interaction grows, and they decrease after a
maximum which depends on $V_0$. On the other hand, the compensated
protocol is perfectly able to reproduce the qualitative trend of the
homogeneous crossover. 

We have shown that the detection of the BCS-BEC crossover in the AHM
by means of cold-atoms in optical lattices is not
straightforward. Using the same trapping potential and increasing the
value of the attractive potential $U$, we are not able to reach a
proper BEC regime because the fermionic cloud collapses into a packed 
``band-insulating'' state with two fermions per site. 
This physics is reflected in the most important observables, including
the local spectral function, the local energy gap and the momentum
distribution function. Interestingly, the energy gap is more
homogeneous than the superfluid order parameter and siginificantly
deviates from local-density approximation. 

The limitations introduced by the trapping potential can
be overcome by tuning the strength of the potential in order to keep
the density at the center of the trap independent on the value of
$U$. This simple choice leads to an essentially fixed density pattern
which allows to recover the main features of the lattice BCS-BEC
crossover. A similar protocol should also be used to study more
complex  situations with population\cite{Dao2008PRL} and/or
mass\cite{Dao2007PRB,Dao2012PRA} imbalance between the two fermionic species
in order to reveal new exotic phases such as Sarma states and FFLO
superfluidity.

This work is financed by ERC/FP7 through the Starting Independent
Grant ``SUPERBAD'',  Grant Agreement No. 240524. We also acknowledge
support of LEMSUPER project (Grant Agreement no. 283214)

\bibliography{localbib}

\end{document}